  \documentclass[journal]{IEEEtranTCOM}
\usepackage{amssymb}
\usepackage{multicol}
\usepackage{subfloat}

\normalsize

\ifCLASSINFOpdf
\else
\fi

\begin{document}
\newtheorem{theorem}{Theorem}
\newtheorem{proposition}{Proposition}
\newtheorem{lemma}{Lemma}
\newtheorem{definition}{Definition}
\newtheorem{corollary}{Corollary}
\newtheorem{remark}{Remark}
\newtheorem{example}{Example}
%
\title{On the Optimum Cyclic Subcode Chains of
$\mathcal{RM}(2,m)^*$ for Increasing Message Length  \thanks{This work was supported by the National Natural Science Foundation of China under Grants 61271222 and 60972033.}}

\author{Xiaogang Liu\thanks{
X. Liu is with the
Computer Science and Engineering Department, Shanghai Jiao Tong University, Shanghai 200240, P. R. China,
e-mail:liuxg0201@163.com.},
Yuan Luo
\thanks{Y. Luo is the corresponding author and with the
Computer Science and Engineering Department,
Shanghai Jiao Tong University, Shanghai 200240, P. R. China, e-mail: yuanluo@sjtu.edu.cn.} and
 Kenneth W. Shum
\thanks{ K. W. Shum is with
Institute of Network Coding, the Chinese University of Hong Kong, Hong Kong, P. R. China, e-mail:
wkshum@inc.cuhk.edu.hk}}

\markboth{IEEE Transactions on Communications}%
{Submitted paper}

\maketitle

\begin{abstract}
The distance profiles of linear block codes can be employed to design variational coding scheme for encoding message with variational length and getting lower decoding error probability by large minimum Hamming distance. 
 Considering convenience for encoding, we focus on the distance profiles with respect to cyclic subcode chains (DPCs) of cyclic codes over $GF(q)$ with length $n$ such that $\mbox{gcd}(n,q) = 1$.
In this paper the optimum DPCs and the corresponding optimum cyclic subcode chains are investigated on the punctured second-order Reed-Muller code $\mathcal{RM}(2,m)^*$ for increasing message length,
where two standards on the optimums are studied according to the rhythm of increase.
\end{abstract}

\begin{IEEEkeywords}
Boolean function,
 distance profile with respect to cyclic subcode chain (DPC),
  exponential sum,
  Reed-Muller code,
  symplectic matrix.
\end{IEEEkeywords}

%
\IEEEpeerreviewmaketitle

\section{Introduction} \label{sec1}
In variational transmission system with linear block code, the changes of the amount of user data will lead to the increase or decrease of the message length, and then lead to the expansion or contraction of linear subcodes.
One example is the transport format combination indicator (TFCI)
in the $3$rd Generation Partnership Project($3$GPP) of CDMA,
which receives about five hundred patents according to the site of US Patent and Trademark Office
(http://patft.uspto.gov/netahtml/PTO/search-adv.htm). Considering convenience for encoding,
we focus on the problem of stepwise expansion of cyclic subcodes while keeping the minimum Hamming distances as large as possible, which is a key parameter for evaluating decoding ability. In this paper,
the distance profiles with respect to cyclic subcode chains (DPCs) are introduced to deal with this problem
on the punctured second-order Reed-Muller code $\mathcal{RM}(2,m)^*$.

The distance profiles and the optimum distance profile (ODP) of a linear block code are about how to select and then include or exclude the basis codewords one by one while keeping the minimum distances of
the generated subcodes as large as possible. The concept was introduced by A. J. Han Vinck and Y. Luo in \cite{HL}, and then investigated for general properties in \cite{HL1} and for a lower bound on the second-order Reed-Muller codes by Y. Chen and A. J. Han Vinck in \cite{CL}. It can be used to get better error correcting ability in channel coding for informed decoders, see M. van Dijk, S. Baggen, and L. Tolhuizen \cite{MG}, and to design the TFCI in CDMA system, see H. Holma and A. Toskala \cite{HT} and R. Tanner and J.Woodard \cite{TW}.

One problem is that, for a given linear block code, the algebraic structure of some subcodes may be lost although the properties of the original code may be good, and vice versa. Here we would like to consider cyclic codes and cyclic subcodes, which imply the convenience of encoding at least. In fact, the successive expansion of cyclic subcodes provide {\bf a cyclic subcode chain}, and the minimum distances of the generated cyclic subcodes form a decreasing distance sequence.

In this paper, we mainly focus on the punctured second-order Reed-Muller code $\mathcal{RM}(2,m)^*$.
The basic knowledge is presented in Section \ref{secII}, which includes distance profile, dimension profile,
dictionary order, inverse dictionary order, Standard I, Standard II and some counting properties of cyclic subcode chains.
In Section \ref{secIII}, the optimum distance profile with respect to cyclic subcode chains under Standard II,
i.e. ODPC-II$^{inv}$, is
studied under one specification that the second  selected cyclic subcode is the punctured first-order Reed-Muller code.
The result of Section \ref{secIII} is suboptimum or a lower bound on ODPC-II$^{inv}$,
but deduces the real ODPC-II$^{inv}$ of $\mathcal{RM}(2,m)^*$
when $m$ is even in Section \ref{secIV}.
Section \ref{secV} is about some optimum distance profiles under certain requirement for most classes in Standard I,
the requirement of which is common.
When $m$ is in the form of a power of $2$, we also get a real optimum one in this section.  Final conclusion is in Section \ref{secVII}.

\section{Preliminaries} \label{secII}


 There are five subsections in this section, which are about the basic definitions of
 distance profile of a linear block code (DPB),
 the optimum distance profile of a linear block code (ODPB),
 distance profile with respect to cyclic subcode chain of a cyclic code (DPC),
 and the optimum DPCs under two respective standards (ODPC-I and ODPC-II), etc.
 In addition, general results about the cyclic subcode chains are presented.

\subsection{Distance Profiles and Subcode Chains of a Linear Block Code\label{secII.I}}

Let $C$ be an $[n, k]$ linear code over $GF(q)$ and denote $C_0=C$.  A sequence of linear subcodes
\[
C_0\supset C_1\supset\cdots\supset C_{k-1}
\]
is called a {\bf subcode chain}, where $\dim[C_i]=k-i$.
An increasing sequence
\[
d[C_0]\le d[C_1]\le\cdots\le d[C_{k-1}]
\]
is called a {\bf distance profile of the linear block code $C$ (DPB)},
where $d[C_i]$ is the minimum Hamming distance of the subcode $C_i$.
It is easy to see that a distance profile
is with respect to a subcode chain.


In the comparison of distance profiles,
the inverse dictionary order is for expanding subcodes, i.e. for increasing the message length, which is on the topic of this paper. In details,
for any two integer sequences of length $k$, $a_0, \ldots, a_{k-1}$ and
$b_0, \ldots, b_{k-1}$, we say that $a_0, \ldots, a_{k-1}$ is
larger than $b_0, \ldots, b_{k-1}$ in the inverse dictionary order if
there is an integer $t$ such that
\begin{eqnarray*}
a_i = b_i \quad\mbox{for $k-1\ge i\ge t+1$}, \ \mbox{and} \ a_t>b_t.
\end{eqnarray*}
We say that $a_0, \ldots, a_{k-1}$ is an upper bound on
$b_0, \ldots, b_{k-1}$ in the inverse dictionary order if
$a_0, \ldots, a_{k-1}$ is larger than or equal to $b_0, \ldots, b_{k-1}$.

A distance profile of an $[n, k]$ linear block code $C$ is called the {\bf optimum distance profile in the inverse dictionary order}, which is denoted by ODPB$^{inv}$:
\[
ODPB[C]^{inv}_0, ODPB[C]^{inv}_1, \ldots, ODPB[C]^{inv}_{k-1},
\]
if it is an upper bound on any distance profile of $C$ in that order.
The ODPB$^{inv}$ will show you how to decrease the minimum distances (a decoding ability) as slowly as possible
when expanding the dimensions of the subcodes one by one in a variational transmission system.
The existence and uniqueness of the optimum distance profile of a linear block code are obvious. A chain that achieves the optimum distance profile is called an {\bf optimum chain} in that order.

\subsection{Distance Profiles with Respect to Cyclic Subcode Chains\label{secII.III}}

Although the properties of some applied linear codes may be good,
it is known that in many cases few algebraic structures are left in its subcodes, and vice verse.
In this paper, we consider the distance profiles with respect to cyclic subcode chains of an $[n, k]$ cyclic code $\mathcal{C}$ over $GF(q)$, where $\mbox{gcd}(n, q)=1$.

 A {\bf cyclic subcode chain} of $\mathcal{C}$ is a chain of cyclic subcodes such that:
\[
\mathcal{C}_{\tau_0}\supset \mathcal{C}_{\tau_1}\supset\cdots\supset \mathcal{C}_{\tau_{\lambda-1}}\supset \{0^n\},
\]
where $\mathcal{C}_{\tau_0}=\mathcal{C}$ and there is no cyclic subcodes between any two neighbors in the chain,
i.e. there does not exist a cyclic code $\mathcal{C}^*$ such that $\mathcal{C}_{\tau_u}\supset \mathcal{C}^*\supset \mathcal{C}_{\tau_{u+1}}$.
The increasing sequence
\[
d[\mathcal{C}_{\tau_0}]\le d[\mathcal{C}_{\tau_1}]\le\cdots\le d[\mathcal{C}_{\tau_{\lambda-1}}]
\]
is called the {\bf distance profile with respect to the cyclic subcode chain (DPC)},
where $\lambda$ is called the length of the profile or the length of the chain. The decreasing sequence
\[
\dim[\mathcal{C}_{\tau_0}]>\dim[\mathcal{C}_{\tau_1}]>\dim[\mathcal{C}_{\tau_2}] > \cdots > \dim[\mathcal{C}_{\tau_{\lambda-1}}]\, \ 
\]
is called the {\bf dimension profile with respect to the cyclic subcode chain}. In general, math calligraphy $\mathcal{C}_i$ denotes an irreducible cyclic code with primitive idempotent $\theta_{l_i}^*$ (Subsection \ref{secIII.I}), and $\mathcal{C}_{\tau_u}$ denotes a cyclic subcode in a chain.

In the comparison among the DPCs in the inverse dictionary order,
according to the dimension profiles or not, two standards are introduced as follows respectively.

\subsubsection{Standard I}

For a given cyclic code $\mathcal{C}$, the lengths of its DPCs are the same, see \cite{LH}.
In order to compare its DPCs,
a classification on the cyclic subcode chains is introduced as follows.
Two chains with length $\lambda$ are set to be in the same class if they have the same dimension profile, i.e.
\[
\dim[\mathcal{C}^1_{\tau_u}]=\dim[\mathcal{C}^2_{\tau_u}]\, \ \mbox{for $0\le u\le \lambda-1$},
\]
where the superscripts 1 and 2 denote the two chains respectively. In each class,
the corresponding DPCs can be compared with each other in the inverse dictionary order, and
we are interested in the optimum one denoted by {\bf ODPC-I$^{inv}$}.
 The corresponding analysis is said to be under {\bf Standard I}.
Some counting properties of the classification are presented in Section \ref{secII.V}.

\subsubsection{Standard II}

For a given cyclic code $\mathcal{C}$,
the distance profiles of any two cyclic subcode chains can be compared directly in the inverse dictionary order, and the analysis without the condition of same dimension profile is said to be under {\bf Standard II}.
The optimum one is denoted by {\bf ODPC-II$^{inv}$}. A cyclic subcode chain that achieves the ODPC (I or II) is called an {\bf optimum cyclic subcode chain} correspondingly.


 Standard I considers dimension profile prior to distance profile, and
Standard II considers distance profile prior to dimension profile.
Since Standard II is without the condition of same dimension profile and Standard I is with the condition, ODPC-II is an upper bound on ODPC-I for each class.
As to Standard I, there are different optimum cyclic subcode chains in different classes, and
the corresponding ODPC-Is can be different.
As to Standard II, there may be more than one optimum cyclic subcode chains, but there exists only one ODPC-II.

\subsection{Key Parameters of Cyclic Subcode Chains} \label{secII.V}

Let $\mathcal{C}$ be an $[n, k]$ cyclic code over $GF(q)$ such that $\mbox{gcd}(n, q)=1$.
Its generator polynomial $g(x)$ is a product of
some distinct minimal polynomials. Let $P$ be the set of the
minimal polynomials that are factors of $g(x)$, and $J(v)$ be the number of the
polynomials with degree $v$ in $P$. Let $A$ be the set of
all minimal polynomials over $GF(q)$ that are factors of $x^n-1$.

Let $m$ be the multiplicative order of $q$ modulo $n$, i.e. $\mbox{ord} (q, n)$, and the integers modulo $n$ are considered in $\{1, 2,\ldots, n\}$.
The $q$-cyclotomic coset modulo $n$ which contains $s$ is $\{s, sq, sq^2, ..., sq^{m_s-1}\}$,
where $m_s$ is the smallest positive integer such that $s=sq^{m_s}$ mod $n$, i.e. $n|s(q^{m_s}-1)$.

\begin{lemma}\label{th1} (Theorem 1, \cite{LH})
For the cyclic code $\mathcal{C}$, we have 
\begin{itemize}
\item
The length of its cyclic subcode chains is $\lambda=|A\setminus P|=\sum_{v: v|m} (L(v)-J(v))$,
where $L(v)$ is the number of $q$-cyclotomic cosets modulo $n$ with size $v$, i.e. $L(v)=\sum_{g\in G(v)} \frac{\varphi(n/g)}{v}$,
$G(v)=\{g: v=\mbox{ord}(q, n/g), g|n\}$ and $\varphi(\cdot)$ is the Euler function.
\item
The number of its cyclic subcode chains is $\lambda!$,
i.e. $\lambda$ factorial.
\item
The number of the chains in each class is $\mu=\prod_{v: v|m} (L(v)-J(v))!$.
\item
The number of classes is $\frac{\lambda!}{\mu}$.
\end{itemize}
\end{lemma}

 \begin{example}
 Assume that $q=2, n=21$, then $m=6$. Let $\mathcal{C}$ be the cyclic code with generator polynomial $g_1(x)=(1+x^2+x^3)(1+x+x^3)=1+x+x^2+x^3+x^4+x^5+x^6 $. Then
\[
\begin{array}{ll}
&J(1)=J(2)=J(6)=0, J(3)=2; \\
&L(1)=1,L(2)=1,L(3)=2,L(6)=2.
\end{array}
\]
From Lemma \ref{th1}, we have
$\lambda = 4, \lambda !=24, \mu =2$ and ${\lambda ! \over {\mu}} = 12$.

In addition, the set $A$ is
\[
\{g_1'(x), g_1(x)/{g_1'(x)}, g_2(x), g_3(x), g_4(x), g_5(x)\}
\]
 where
$g_1'(x)=1+x^2+x^3, g_2(x)=1+x+x^2 +x^4 +x^6 $, $g_3(x) = 1 +x^2 +x^4+x^5+x^6 $, $g_4(x)=1+x+x^2 $, and $g_5(x)=1+x$.
In the investigation of Standard I, for the class of dimension profile $15, 9, 8, 6$, there are $\mu =2$ cyclic subcode chains.
One such chain can be obtained from the cyclic subcodes generated by the following polynomials respectively:
 \begin{eqnarray*}
 g_1(x), \  g_1(x)g_2(x),\  g_1(x)g_2(x)g_5(x), \  g_1(x)g_2(x)g_5(x)g_4(x).
\end{eqnarray*}
Using Matlab, we find that the corresponding DPC is $2, 6, 6, 8$.
In fact, for this class of dimension profile, the two cyclic subcode chains have the same DPC, that is the  ODPC-I$^{inv}$ is $d_{\tau_0}=2,\, d_{\tau_1}=6,\, d_{\tau_2}=6,\, d_{\tau_3}=8$.
\end{example}

\section{Suboptimums with respect to ODPC-II$^{inv}$ of $\mathcal{RM}(2,m)^*$} \label{secIII}

Let $\mathcal{RM}(2,m)$ be the second-order Reed-Muller code.
Deleting the first coordinate of each codeword, the well known punctured code $\mathcal{RM}(2,m)^*$ is obtained,
which is a cyclic code of length $n = 2^m-1$ and dimension
$k = 1 + \left(\begin{array}{c}
m\\
1 \end{array}\right) +  \left(\begin{array}{c}
m\\
2 \end{array}\right)$.
 The dual of the Reed-Muller code $\mathcal{RM}(r,m)$ is $\mathcal{RM}(m-r-1,m)$, which is the extended Hamming code when $r=1$; $\mathcal{RM}(r,m)$ itself is a subcode of the extended BCH code of designed distance $2^{m-r}-1$. Reed-Muller codes are widely used, for example,
in some communication systems which require fast decoding, in the localization of Malicious Nodes \cite{KW}, in the Power Control of OFDM Modulation \cite{DJ}, \cite{KGP}, and so on. In \cite{CL}, linear block codes with optimum distance profiles
(ODPB), as defined in \cite{HL}, were investigated, and the authors provided a lower bound on the optimum distance profile of the second-order Reed-Muller codes, which is proved to be tight for $m\leq 7$.

\subsection{The Case of $m = 2t+1$ in $\mathcal{RM}(2,m)^*$\label{secIII.I}}

In this subsection Lemma \ref{th2} is cited to show the
weight distributions of some subcodes of $\mathcal{RM}(2,m)$. Then, by using the symplectic forms derived from the codewords of $\mathcal{RM}(2,m)$ \cite{MS},
we show how to construct the optimum distance profile under certain requirement with respect to ODPC-II$^{inv}$ in Theorem \ref{th4},
where the profile is calculated in Lemma \ref{pro}.

Let $\alpha$ be a primitive $n$th root of unity in $GF(q)$, where $q=2^m$ and $n=2^m-1$ is the length of the cyclic code $\mathcal{RM}(2,m)^*$.
Let $\mathcal{D}_s$ be the cyclotomic coset containing $s$, with primitive idempotent denoted by $\theta_s$, and use $\theta^*_s$ to denote the primitive idempotent corresponding to $\mathcal{D}_{-s}$. Easy to see that $\mathcal{D}_s$ and $\mathcal{D}_{-s}$ have the same size. The nonzeros of $\theta_s$ and $\theta^*_s$ are $\{\alpha^i:i\in \mathcal{D}_s\}$ and $\{\alpha^i:i\in \mathcal{D}_{-s}\}$ respectively.
Note that, $\theta_0$, $\theta_1^*$, $\theta_{l_i}^* (1\leq i \leq t)$ are all the primitive idempotents contained in $\mathcal{RM}(2,m)^*$ , which correspond to all the minimal cyclic subcodes of $\mathcal{RM}(2,m)^*$.
Any cyclic subcode of $\mathcal{RM}(2,m)^*$ can be given by idempotent of the form
\[
a_{-1}\theta_0+a_0\theta_1^*+\sum_{j=1}^{t}a_j\theta_{l_j}^*,  \quad a_j\in \{0, 1\}, -1\leq j\leq t.
\]


\begin{lemma}\label{th2}(Ch.15, \cite{MS}.)
Let m=2t+1, and let h be any number in the range $1\le h \le t$. Then there exists a
\[
[2^m, m(t-h+2)+1, 2^{m-1}-2^{m-h-1}]
\]
subcode $\mathcal{RM}_{2t+1}^h$ of $\mathcal{RM}(2,m)$. It is obtained by extending the cyclic subcode of $\mathcal{RM}(2,m)^*$ having idempotent
\[
\theta_0+\theta_1^*+\sum_{j=h}^{t}\theta_{l_j}^*,  \quad l_j=1+2^j.
\]
The code has codewords of weights $2^{m-1}$ and $2^{m-1}\pm2^{m-h'-1}$ for all $h'$ in the range $h\le h' \le t$.
\end{lemma}

\begin{remark}
One cyclic subcode chain of $\mathcal{RM}(2,m)^*$ can be obtained from Lemma \ref{th2}. But there are many other chains. We will compare and select the suboptimum one in Theorem \ref{th4}.
\end{remark}

\begin{lemma} \label{even2}(pp. 453, \cite{MS})
Let $ \Phi_h $ be the set of symplectic forms derived from the codewords of the second-order Reed-Muller code $\mathcal{RM}(2,m)$. Suppose that it has the property that the rank of every nonzero form in $ \Phi_h $ is at least $2h$, and the rank of the sum of any two distinct forms in $\Phi_h$ is also at least $2h$, here $h$ is some fixed number in the range $1\leq h\leq \lfloor m \rfloor $, then the maximum size of such a set $ \Phi_h$ is $2^{(2t+1)(t-h+1)}$ if $m=2t+1$, and $2^{(2t+1)(t-h+2)}$ if $m=2t+2$.
\end{lemma}

For each $i (1\leq i \leq t)$, let $ \mathcal{RM}_{2t+1}^{i}$ denote the subcode of $\mathcal{RM}(2,m)$ suggested in Lemma \ref{th2}, and $\mathcal{RM}_{2t+1}^{i*}$ denote the cyclic subcode obtained from $\mathcal{RM}_{2t+1}^{i}$ by puncturing the first coordinate.
Express the one-dimensional cyclic subcode corresponding to $\theta_0$ by $\mathcal{RM}(0,m)^*$, 
 and the punctured first-order Reed-Muller code by $\mathcal{RM}(1,m)^*$. 

\begin{corollary}\label{MNEC}
 Lemma \ref{th2} implies the following cyclic subcode family with a nested structure
\[
 \mathcal{RM}(0,m)^* \subset \mathcal{RM}(1,m)^* \subset \mathcal{RM}_{2t+1}^{t*}  \subset \cdots \subset \mathcal{RM}_{2t+1}^{1*}. 
\]
\end{corollary}

The following result concerns the minimum distances of the cyclic subcodes in Corollary \ref{MNEC}.
\begin{lemma} \label{pro}
The distance profile of the cyclic subcode chain given in Corollary \ref{MNEC} is
\begin{eqnarray*}
&  d_{\tau_u}=2^{2t}-2^{2t-u-1}-1 (0\leq u\leq t-1), \\
& d_{\tau_{t}} = 2^{m -1 }-1 =  2^{2t}-1,  d_{\tau_{t+1}} =  2^{m }-1 = 2^{2t+1}-1.
\end{eqnarray*}
\end{lemma}

\begin{IEEEproof}
Let $c$ be a codeword of the cyclic subcode $\mathcal{RM}_{2t+1}^{i*} (1\le i\le t )$ in Corollary \ref{MNEC}, with symplectic form of rank $2d (1\le d\le t)$.
From Theorem 4 and Theorem 5 in Chapt15 \cite{MS}, rewrite the Boolean function $f$ as $T(y)=\sum_{i=1}^{d}y_{2i-1}y_{2i}+L(y)+\epsilon$,
here $L(y)$ and $\epsilon$ are arbitrary.
Choose $T(y)=\sum_{i=1}^{d}y_{2i-1}y_{2i}+y_1+y_2+1$.
As in the proof of Theorem 5, Chapt15 \cite{MS}, the final expression is nonzero for $2^{m-1}-2^{m-d-1}$ coordinates.
Since $T(y)=1$ when $y_i=0,1\leq i\leq m$, deleting the first coordinate, the corresponding weight is $2^{m-1}-2^{m-d-1}-1$.
\end{IEEEproof}

\begin{example}
For $t=2$, that is $m=5, n=31$. Lemma \ref{pro} provides a distance profile $d_{\tau_0}=7,\, d_{\tau_1}=11,\, d_{\tau_2}=15,\, d_{\tau_3}=31$ with dimension profile $16,\, 11,\, 6,\, 1$. The generated nontrivial cyclic codes [31,11,11] and [31, 6, 15] are optimal \cite{G}.
\end{example}

\begin{theorem}\label{th4}
Let $m=2t+1$ where $t\geq 1$. Then for the punctured second-order Reed-Muller code $\mathcal{RM}(2,m)^*$,
if we are requiring that, the second selected cyclic subcode is $\mathcal{RM}(1,m)^*$,
the distance profile of the cyclic subcode chain in Corollary \ref{MNEC} is optimum  under Standard II.
\end{theorem}

\begin{IEEEproof}
To get the optimum distance profile under the requirement, we have to add the primitive idempotents one by one
accumulatively from $\theta_0, \theta_1^*, \theta_{l_j}^* (1\leq j\leq t)$, and at the same time try to make the minimum distance of the cyclic subcode generated by the accumulative sum as large as possible according to the following steps.

1)\ \ It is obvious that the first cyclic subcode must be $\mathcal{RM}(0,m)^* $ which has
the largest minimum distance $d_{\tau_{t+1}}=2^m-1=n$ (code length).
That is to say $\theta_0$ is selected in this step.
Then all the cyclic subcodes of the chain are
self-complementary.

2)\ \ In the requirement, the second cyclic subcode has idempotent $\theta_0+\theta_1^*$
which generates the punctured first-order Reed-Muller code $\mathcal{RM}(1,m)^*$ satisfying $d_{\tau_{t}}=2^{m-1}-1$.
Briefly $\theta_1^*$ is selected here.

3)\ \ With decreasing $h$ from $t$ to $1$ and selecting $l_{j_s}$, idempotents $\theta_0$, $\theta_0+\theta_1^*$, $\theta_0+\theta_1^*+\sum_{s=h}^{t}\theta_{l_{j_s}}^* $ where $1\le j_s \le t$,
can provide any cyclic subcode chain with beginner $\mathcal{RM}(0,m)^*$ and $\mathcal{RM}(1,m)^*$.
In the ${(t-h+3)}$th step,
the generated cyclic subcode is denoted by $\mathcal{C}_{\tau_{h-1}}$.
Note that, in the ${(t-h+3)}$th step of Corollary \ref{MNEC}, the cyclic subcode $ \mathcal{RM}_{2t+1}^{h*}$ has minimum distance $d_{\tau_{h-1}}= 2^{m-1}-2^{m-h-1}-1$ using Lemma \ref{pro}.

The set of symplectic forms contained in $\mathcal{C}_{\tau_{h-1}}$ is with size $N = 2^{m(t-h+1)}$.
According to Lemma \ref{even2}, the maximum size of the set of symplectic forms satisfying that each element has rank $\ge 2{(h+1)}$ is at most $2^{(2t+1)(t-(h+1)+1)}$, which is smaller than $N$. So there must be some symplectic forms in $\mathcal{C}_{\tau_{h-1}}$ which have ranks $2d < 2{(h+1)}$. According to the proof Lemma \ref{pro}, $\mathcal{C}_{\tau_{h-1}}$ has a codeword of weight $2^{m-1}-2^{m-d-1}-1\le 2^{m-1}-2^{m-h-1}-1=d_{\tau_{h-1}}$.

In one word, the distance profile 
$d_{\tau_{0}} \le d_{\tau_{1}} \le \cdots \le d_{\tau_{t}} \le d_{\tau_{t+1}} $
of the cyclic subcode chain given in corollary \ref{MNEC} is optimum under the requirement of the theorem.
\end{IEEEproof}

\subsection{The Case of $m = 2t +2 $ in $\mathcal{RM}(2,m)^*$ } \label{secIII.II}

In this subsection the punctured second-order Reed-Muller code $\mathcal{RM}(2,m)^*$ is studied for  the case of $m=2t+2$ , where $t\geq 1$ is a positive integer.
This case is in parallel with the case of $m=2t+1$.
Corresponding to Lemma \ref{th2} for Subsection \ref{secIII.I}, Lemmas \ref{EC} is stated for the weight distributions of certain subcodes of $\mathcal{RM}(2,m)$ . Just like Theorem \ref{th4}, a suboptimum distance profile with respect to ODPC-II$^{inv}$ is presented in Theorem \ref{even22} .

\begin{lemma} \label{EC}(Theorem 3.6, \cite{CL}.)
 Let $m=2t+2$, and let $h$ be any number in the range $1\leq h\leq t+1$. Then there exists a
 \[ 
[2^m,m(t-h+2)+m/2+1,2^{m-1}-2^{m-h-1}]
\]
subcode $\mathcal{RM}_{2t+2}^h$ of $\mathcal{RM}(2,m)$. It is obtained by extending the cyclic subcode of $\mathcal{RM}(2,m)^*$ having idempotent
\[ 
\theta _0+\theta _1^*+\sum_{j=h}^{t+1}\theta _{l_j}^*,\quad l_j=1+2^j.
\]
The code has codewords of weights $2^{m-1}$ and $2^{m-1}\pm 2^{m-h'-1}$ for all $h'$ in the range $h\leq h'\leq t+1$.
\end{lemma}

Define $\mathcal{RM}{(0,m)}^*$, $\mathcal{RM}{(1,m)}^*$ and $\mathcal{RM}_{2t+2}^{u*} (1\leq u \leq t+1)$
 analogously as in  Subsection \ref{secIII.I}. 

\begin{corollary} \label{evenc1} 
 Lemma \ref{EC} implies the following cyclic subcode family with a nested structure
\begin{equation}\label{OPECC}
   \mathcal{RM}{(0,m)}^*\subset \mathcal{RM}{(1,m)}^*\subset \mathcal{RM}_{2t+2}^{(t+1)*}\subset \cdots \subset \mathcal{RM}_{2t+2}^{1*}.
\end{equation}
\end{corollary}


Similar to Lemma \ref{pro}, the following result is for the case of $m=2t+2$.
\begin{lemma} \label{pro2}
The distance profile of the cyclic subcode chain given in Corollary \ref{evenc1} is
\begin{eqnarray*}
&   d_{\tau_u}=2^{2t+1}-2^{2t-u}-1 (0\leq u\leq t),  \\
& d_{\tau_{t+1}} = 2^{m-1} -1 =2^{2t+1}-1,  d_{\tau_{t+2}} =  2^{m}-1 = 2^{2t+2}-1. 
\end{eqnarray*}

\end{lemma}

\begin{example}
For $t=1$, that is $m=4, n=15$. Lemma \ref{pro2} provides a distance profile $d_{\tau_0}=3,\, d_{\tau_1}=5,\, d_{\tau_2}=7,\, d_{\tau_3}=15$ with dimension profile $11,\, 7,\, 5,\, 1$. The generated nontrivial cyclic codes [15,11,3], [15, 7, 5] and [15, 5, 7] are optimal \cite{G}.
\end{example}

Using Lemma \ref{EC}, Theorem \ref{even22} can be verified which is the counterpart of Theorem \ref{th4}.

\begin{theorem} \label{even22}
Let $m=2t+2$ where $t\geq 1$. Then for the punctured second-order Reed-Muller code $\mathcal{RM}(2,m)^*$, if we are requiring that, the second selected cyclic subcode is $\mathcal{RM}(1,m)^*$, the distance profile of the cyclic subcode chain in Corollary \ref{evenc1} is optimum under Standard II.
\end{theorem}

\section{The ODPC-II$^{inv}$ of $\mathcal{RM}(2,m)^*$ when $m=2t+2$}  \label{secIV}

In this section we investigate the exact ODPC-II$^{inv}$ of the punctured second-order Reed-Muller code $\mathcal{RM}(2,m)^*$ when $m=2t+2$.
Subsection \ref{secIV.I} is about the basic background on cyclic codes. In Subsection \ref{secIV.II},  the weight distributions of cyclic subcodes of $\mathcal{RM}(2,m)^*$ are studied, and then Theorem \ref{even22} is reinvestigated in Theorem \ref{even222}.

\subsection{Basic Results on the Weight of Codeword in Cyclic Codes}\label{secIV.I}
Many of the following preliminaries are well referred to \cite{FL,KA2,KA1,RH,LTW,VD,WO}
which list the properties on weight distributions of cyclic codes, trace functions, exponential sums, quadratic forms and their relations. See also
\cite{DHL,DHM,H}
for binary sequences.

 Let $q=2^m$ and F$_q$ be the finite field of order $q$. Let $\pi$  be a primitive element of F$_q$, \mbox{Tr}:F$_{2^m}\to \mbox{F}_2$ be the trace mapping, and $e(x)=(-1)^{\mbox{\mbox{Tr}}(x)}$ is the canonical additive character on F$_q$.
 For the binary cyclic code $\mathcal{C}$ with length $l=q-1$ and nonzeros $\pi^{-s_{\lambda}}$, $1\leq s_{\lambda} \leq q-2(1\leq {\lambda}\leq u)$, the codewords in $\mathcal{C}$ can be expressed by
\[
c(\alpha_1,\ldots,\alpha_u)=(c_0,c_1,\ldots,c_{l-1})\quad(\alpha_1,\ldots,\alpha_u\in \mbox{F}_q)
\]
where $c_i=\sum\limits_{\lambda=1}^u\mbox{\mbox{Tr}}(\alpha_{\lambda}\pi^{is_{\lambda}})(0\leq i\leq l-1)$. Therefore the Hamming weight of the codeword $c=c(\alpha_1,\ldots,\alpha_u)$ is
\begin{eqnarray}
w_H(c)
 & = & l-\#\{i|0\leq i\leq l-1,c_i=0\} \nonumber\\
 & = & l-{l\over 2}-{1\over 2}\sum\limits_{x\in \mbox{F}_q^*}(-1)^{\mbox{\mbox{Tr}}(f(x))} \nonumber \\
& = & 2^{m-1}-{1\over 2}S(f,m) \label{wt}
\end{eqnarray}
where $f(x)=\alpha_1x^{s_1}+\alpha_2x^{s_2}+\cdots +\alpha_ux^{s_u}\in \mbox{F}_q[x]$, and $S(f,m)=\sum\limits_{x\in \mbox{F}_q}e(f(x))$.

For $f(x)=\alpha x^{2^i+1}+\beta x^{2^j+1}+\cdots +\gamma x^{2^k+1} \in \mbox{F}_{q}[x]$, we have $S(f,m)=\sum\limits_{X\in \mbox{F}_p^m}(-1)^{XF_{\alpha,\beta,\ldots,\gamma}X^{T}}$
where $H_{\alpha,\beta,\ldots,\gamma}$ is the matrix of the quadratic form $F_{\alpha,\beta,\ldots,\gamma} (F_{\alpha,\beta,\ldots,\gamma}(X)=\mbox{Tr}(f(x)))$.
  $S(f,m)$ is also denoted by $T{(\alpha,\beta,\ldots, \gamma)}$.
For a quadratic form $F$ with corresponding matrix $H$, define $r_F$ to be the rank of the skew-symmetric matrix $H+H^T $. Then $r_F$ is even.

Let $f(x)= \alpha x^{2^i+1}$ and $\mbox{\mbox{Tr}}(f(x))= XH_{\alpha}X^T$ where $\alpha \in  \mbox{F}_q^*$. The following lemma can be deduced by using a similar argument as in \cite{FL,LTW}.

 \begin{lemma}\label{LTWLC}
For $ \alpha \in \mbox{\rm{F}}_q  \backslash \{0\}  $, let $r_ {\alpha }$ be the rank of $H_{\alpha }+H_{\alpha }^T$. Then $r_{\alpha}=m$ or $m-\mbox{gcd}(2i,m)$ where $1\le i \le t$.
\end{lemma}

\subsection{Main Results} \label{secIV.II}

Now we focus on the ODPC-II$^{inv}$ of $\mathcal{RM}(2,m)^*$ when $m$ is even.
Lemma \ref{PRIMC}, Lemma \ref{EHM}, and Corollary \ref{EE22} investigate the existence of certain one-weight minimal cyclic code.
Then in Subsection \ref{secIV.II.I}, Corollary \ref{EC222} gives the optimum cyclic subcode chain for the case of $m=2^s$. In Subsection \ref{secIV.II.II}, Corollary \ref{ECC222} considers the case when $m=2t+2$ is not a power of $2$. Final results are given by Theorem \ref{even222} 
with an example.

In the subsequent, Lemma \ref{XHG} is about the greatest common divisor of $2^{\alpha}+1$ and $2^{\beta}-1$; Lemma \ref{CCL} is about the size of cyclotomic cosets $\mathcal{D}_{l_i}$; Lemma \ref{LTWLS} is about the exponential sums of quadratic forms. They will be used in Lemma \ref{PRIMC} about the existence of certain one-weight minimal cyclic code and then support the determination
of the optimum distance profile. In addition, define the $2$-adic order function $\nu_2(*)$, such that $\mbox{$\nu_2(n)=s$ for $n=2^sn'$ where $n'$ is odd}$.

\begin{lemma}(Lemma 5.3, \cite{XHD})\label{XHG}
Let $\alpha,\beta \geq 1 $ be integers. Then
\[ 
\mbox{gcd}(2^{\alpha}+1,2^{\beta}-1) =
 \left\{
\begin{array}{cc}
2^{\mbox{gcd}(\alpha,\beta)}+1 & \textrm{if} \ \nu_2(\beta ) > \nu_2(\alpha)  , \\
1 & otherwise.
\end{array}
 \right.
\]
\end{lemma}

\begin{lemma}(Lemma B.2, \cite{CL})\label{CCL}
If $m=2t+1$ is odd, then for $l_i=1+2^i$, the cyclotomic coset $\mathcal{D}_{l_i}$ has size
\[
|\mathcal{D}_{l_i}|=m, \ 1 \leq i \leq t.
\]
If $m=2t+2$ is even, then for $l_i=1+2^i$, the cyclotomic coset $\mathcal{D}_{l_i}$ has size
\[ 
|\mathcal{D}_{l_i}| =
\left\{
\begin{array}{cc}
 m, & 1\leq i\leq t    \\
m/2, & i=t+1.
\end{array}
\right.
\]
\end{lemma}

\begin{lemma}(Lemma 1, \cite{LTW})\label{LTWLS}
For the quadratic form $F(X)=XHX^T$ defined as before,
\[
S(f,m)=\sum\limits_{x\in \mbox{F}_2^m}(-1)^{\mbox{Tr}(f(x))} =\sum\limits_{X\in \mbox{F}_2^m}(-1)^{ F(X)} =\pm 2^{m-{r_F\over 2} } \ \mbox{or} \ 0
\]
Moreover, if $r_F=m$, then
\[
S(f,m)=\sum\limits_{X\in \mbox{F}_2^m}(-1)^{ F(X) }=\pm 2^{m\over 2}.
\]
\end{lemma}

For the irreducible cyclic codes $\mathcal{C}_{i} (\theta_{l_i}^*, 1\leq i\leq t)$, the following lemma will be used in Lemma \ref{PRIMC} to characterize their weights.
\begin{lemma}(Corollary 3.7 in Ch.3, \cite{RH})\label{NSFF}
If $e_1$ and $e_2$ are positive integers, then the greatest common divisor of $x^{e_1}-1$ and $x^{e_2}-1$ in $F_q[x]$ is $x^d-1$, where $d$ is the greatest common divisor of $e_1$ and $e_2$. If $e_2=q-1$ and $q=2^m$, then for any positive integer $e_1$, the number of different solutions of $x^{e_1}-1=0$ in $F_q$ is $d$.

\end{lemma}

\begin{lemma} \label{PRIMC}
The irreducible cyclic code $\mathcal{C}_{ i}$ is a one-weight cyclic code if and only if $\mbox{\mbox{gcd}}(2^i+1,2^m-1)=1$, and in this case the only nonzero weight is $2^{m-1}$.
\end{lemma}
\begin{IEEEproof}
 From Lemma \ref{CCL}, the minimal cyclic code $\mathcal{C}_{ i}$ has dimension $m$. By Lemma \ref{LTWLC}, the only possible rank of the corresponding skew-symmetric matrix $H_{\alpha}+H_{\alpha}^T$
is $m$ or $m-\mbox{gcd}(2i,m)$. Applying (\ref{wt}), the weight of the corresponding codeword is
$w_H(c)=2^{m-1}-{1\over 2}T(\alpha)$.

Let $2i'=\mbox{gcd}(2i,m)$ since $m$ is even. According to the possible values of $T(\alpha)$ given by Lemma \ref{LTWLS},
for $\varepsilon =\pm 1$, denote
$ N_{\varepsilon,i}=\big\{\alpha \in \mbox{F}_q\backslash \{0\}|  T(\alpha)=\varepsilon 2^{{m+2i'}\over 2}\big\}, n_{\varepsilon,i} =|N_{\varepsilon,i}|; N_{\varepsilon,0}=\big\{\alpha \in \mbox{F}_q\backslash \{0\}|T(\alpha)=\varepsilon 2^{{m}\over 2}\big\}, n_{\varepsilon,0}=|N_{\varepsilon,0}|; N_0=\big\{\alpha \in \mbox{F}_q\backslash \{0\}|  T(\alpha)=0\big\}, n_0=|N_0|$.

``Only if" part.\quad  Assume $\mathcal{C}_{ i}$ has only one nonzero weight.
 Let $A_i$ and $A_i'$ ($i=0,1,\ldots,2^m-1$) be the weight distributions of $\mathcal{C}_{i}$ and its dual $\mathcal{C}_{ i}^{\perp}$ respectively. 
  It is easy to see that $A_1'=0$. From the MacWilliams identities, see \cite[pp.131]{MS},
\begin{equation}\label{MACIDD}
\sum\limits_{i=1}^n{{iA_i}\over {2^k}}={1\over 2}(n-A_1')={1\over 2}n \  \mbox{in our case},
\end{equation}
here $n= 2^m-1$ is the length of the code, and $k=m$ is the dimension.
Equation (\ref{MACIDD}) implies that if there is only one nonzero weight $j, 1\leq j\leq 2^m-1$, then $A_j = 2^{m} -1$ and $j=2^{m-1}$.
We have $ n_{1,0}= n_{-1,0}= n_{1,i}= n_{-1,i}=0$, and
\begin{equation}\label{EST2}
\begin{array}{lll}
   \sum\limits_{\alpha \in \mbox{F}_q}T(\alpha)^2
     & =& T(0)^2+(2^{m\over 2})^2n_{1,0}+(-2^{m\over 2})^2n_{-1,0}  \\
     & +&(2^{{m+2i'}\over 2})^2n_{1,i}+(-2^{{m+2i'}\over 2})^2n_{-1,i}\\
     &= &2^{2m}+2^m(n_{1,0}+n_{-1,0})\\
     &+&2^{m+2i'}(n_{1,i}+n_{-1,i}) = 2^{2m}
     \end{array}
\end{equation}
where $T(0)=T(\alpha=0)=2^m$.

Now, equation (\ref{EST2}) can be calculated in another way
\begin{equation}\label{EST}
\begin{array}{lll}
  \sum\limits_{\alpha \in \mbox{F}_q}T(\alpha)^2
 &=&\sum\limits_{\alpha \in \mbox{F}_q}\sum\limits_{x,y\in \mbox{F}_q}(-1)^{\mbox{\mbox{Tr}}\left(\alpha \left(x^{2^i+1}+y^{2^i+1}\right)\right)}\\
& =& \sum\limits_{x,y \in \mbox{F}_q}\sum\limits_{\alpha \in \mbox{F}_q}(-1)^{\mbox{\mbox{Tr}}\left(\alpha \left(x^{2^i+1}+y^{2^i+1}\right)\right)} \\
  &= &\sum\limits_{\stackrel{\alpha \in \mbox{F}_q}{x^{2^i+1}+y^{2^i+1}=0}}(-1)^{\mbox{\mbox{Tr}}\left(\alpha \left(x^{2^i+1}+y^{2^i+1}\right)\right)} \\
   &= &2^m \cdot M_2,
   \end{array}
\end{equation}
where $M_2$ is the number of solutions to the equation $x^{2^i+1}+y^{2^i+1}=0$.
From Lemma \ref{NSFF}, easy to find that $M_2=1+(2^m-1)\mbox{\mbox{gcd}}(2^i+1,2^m-1)$.
Thus $M_2=2^m$,
which implies $\mbox{\mbox{gcd}}(2^i+1,2^m-1)=1$.

``If" part.\quad Assume $\mbox{\mbox{gcd}}(2^i+1,2^m-1)=1$. Similar to equations (\ref{EST2}) and (\ref{EST}), we have
\[
\left\{
\begin{array}{lll}
 \sum\limits_{\alpha \in \mbox{F}_q}T(\alpha)
 & = &2^m+2^{m\over 2}n_{1,0}-2^{m\over 2}n_{-1,0} \\
  &+&  2^{{m+2i'}\over 2}n_{1,i}-2^{{m+2i'}\over 2}n_{-1,i}=2^m,  \\
     \sum\limits_{\alpha \in\mbox{F}_q}T(\alpha)^3
 &= &2^{3m}+2^{{3m}\over 2}n_{1,0}-2^{{3m}\over 2}n_{-1,0}\\
 &+& 2^{{3(m+2i')}\over 2}n_{1,i}-2^{{3(m+2i')}\over 2}n_{-1,i}=2^{3m},  \\
   \sum\limits_{\alpha \in \mbox{F}_q}T(\alpha)^4
   &=& 2^{4m}+2^{2m}(n_{1,0}+n_{-1,0})\\
   &+&2^{2m+4i' }(n_{1,i}+n_{-1,i})=2^{4m}.
\end{array}
\right.
\]
Combining with $n_{1,0}+ n_{-1,0}+ n_{1,i}+ n_{-1,i}+n_0=2^m-1$, it is not difficult to see that $n_{1,0}=n_{-1,0}=n_{1,i}=n_{-1,i}=0$ and $n_0=2^m-1$. That is the only nonzero weight is $2^{m-1}$.
\end{IEEEproof}

\begin{lemma}\label{EHM}
 The minimal cyclic code $\mathcal{C}_{m\over 2}$ with primitive idempotent $\theta_{m\over 2}^*$ has dimension $m\over 2$, and only one nonzero weight $2^{m-1}+2^{{m\over 2}-1}$.
\end{lemma}

\begin{remark}
In Lemma \ref{EHM}, $\mbox{\mbox{gcd}}(2^{m\over 2}+1,2^m-1) = 2^{m\over 2}+1 \not=1 $, but the minimal cyclic code $\mathcal{C}_{m\over 2}$ is still a one-weight code. 
\end{remark}

The following lemma will be used in Corollary \ref{EE22} for the nonexistence of certain one-weight irreducible cyclic code.
\begin{lemma} \label{EC22}
Let $m = 2t+2$ where $t\geq 1$. Then $m$ is not a power of $2$ if and only if there exists $1\leq i\leq t$ such that $\mbox{gcd}(2^i+1,2^m-1) = 1$.
\end{lemma}
\begin{IEEEproof}
``Only if" part. Let $m = 2^{u} \cdot m'$ where $m'\ge 3$ is odd and $u\geq 1$. Set $i = 2^{u}$. Then $t = {{m-2} \over 2}$, and $i\leq t$. In Lemma \ref{XHG}, $\nu_{2}(i) = \nu_{2}(m) = u$, so $\mbox{\mbox{gcd}}(2^i+1,2^m-1) = 1$.

``If" part. If $m = 2^s$ is a power of $2$ where $s\geq 2$, from Lemma \ref{XHG} for any positive integer $1\leq i\leq t$, we have $\nu_{2}(i) <\nu_{2}(m)=s$. That is $\mbox{\mbox{gcd}}(2^i+1,2^m-1) = 2^{\mbox{\mbox{gcd}}(i,m)}+1\geq 3$.
\end{IEEEproof}

\begin{corollary}\label{EE22}
Let $m=2^s = 2t+2$ where $s\geq 2$. For  $1\leq i\leq t$, $\mathcal{C}_{ i}$ is not a one-weight cyclic code, and has weights of the forms $2^{m-1}+\varepsilon 2^a$ and $2^{m-1}-\varepsilon 2^{a'}$ where $a, a'$ are positive integers. 
\end{corollary}

\subsubsection{The Optimum Profile when $m$ Is a Power of $2$} \label{secIV.II.I}


The following lemma can be derived from Lemma \ref{EHM} and Corollary \ref{EE22}.
\begin{lemma} \label{XHP}
Let $m=2^s=2t+2$ where $s\geq 2$. For $1\leq i \leq t+1$, the cyclic code $\mathcal{C}_{i}'$ with idempotent $\theta_0 + \theta_{l_i}^*$, has minimum distance less than $2^{m-1}-1$.
\end{lemma}

\begin{corollary}\label{EC222}
Let $m=2^s$ where $s\geq 2$. In the process of selecting the optimum cyclic subcode chain of $\mathcal{RM}(2,m)^*$ under standard $II$, $\theta_0$ and $ \theta_1^*$ will be the first two selected primitive idempotents. Then (\ref{OPECC}) is an optimum cyclic subcode chain.
\end{corollary}
\begin{IEEEproof}
Since the cyclic code $\mathcal{C}_0'$ with idempotent $\theta_0+\theta_1^*$ has minimum distance $2^{m-1}-1$, from Lemma \ref{XHP} 
$\theta_1^*$ should be the second selected primitive idempotent. And the result follows from Theorem \ref{even22}. 
\end{IEEEproof}

\subsubsection{ Other Cases of $m$} \label{secIV.II.II}

In this subsection, Corollary \ref{ECC222} investigates the ODPC-II$^{inv}$ of $\mathcal{RM}(2,m)^*$, where $m=2t+2$ is not a power of $2$.  In fact, Corollary \ref{ECC222} is supported by Lemma \ref{ccc222} and Lemma \ref{EM222} in the investigation of weight distributions and minimum distances.
The cyclic code $\mathcal{C}_{i,j}$ is with idempotent $\theta_{l_i}^*+\theta_{l_j}^*$ and length $2^m-1$, where $1\leq i\not= j\leq t+1$. 
Assume that at least one of $i,j$, let's say $i$, satisfies $\mbox{\mbox{gcd}}(2^i+1,2^m-1) = 1$. 

For the following lemma, we fix some notations. Let ${n_1}$ be an even integer, ${m_1}={n_1}/2$ and $q_1=2^{n_1}$. Let ${k_1}$ be a positive integer, $1\leq k_1 \leq {n_1}-1$ and $k_1\not={m_1}$. Let $d_1=\mbox{gcd}({m_1},k_1)$ and ${d_1'}=\mbox{gcd}({m_1}+{k_1},2{k_1})$. For $\alpha \in \mathbb{F}_{2^{m_1}}, \beta \in \mathbb{F}_{2^{n_1}}$, set $T(\alpha,\beta)=\sum\limits_{x\in \mathbb{F}_q}(-1)^{\mbox{Tr}_1^{m_1}(\alpha x^{2^{m_1}+1})+\mbox{Tr}_1^{{n_1}}(\beta x^{2^{k_1}+1})}$, and $\mathcal{C}'$ is the binary cyclic code of length $l_1=q_1-1$ with nonzeros $\pi^{-(2^{k_1}+1)}$ and $\pi^{-(2^{m_1}+1)}$.

 \begin{lemma}(Theorem 1, \cite{LTW}) \label{CCC221}
  The value distribution of the multi-set $\{T(\alpha,\beta)|\alpha \in \mathbb{F}_{2^{m_1}},\beta\in \mathbb{F}_{q_1}\}$ and the weight distribution of $\mathcal{C}'$ are shown as following 
\begin{enumerate}
\renewcommand{\labelenumi}{$($\mbox{\roman{enumi}}$)$}
\item For the case $d_1'=d_1$,

 \begin{table}[htbp]
 \renewcommand\thetable{\Roman{table}}
 \begin{center}
 \begin{tabular}{|l|l|l|}
\hline
$value$  &   $weight$        & $multiplicity$   \\
\hline
$2^{m_1}$   &   $2^{{n_1}-1}-2^{{m_1}-1}$        & ${{2^{d_1-1}(2^{m_1}-1)(2^{n_1}+2^{{m_1}+1}+1)}\over {2^{d_1}+1}}$   \\
\hline
 $-2^{m_1}$ &   $2^{{n_1}-1}+2^{{m_1}-1}$       & ${{2^{d_1-1}(2^{m_1}-1)(2^{n_1}-2^{{n_1}-{d_1}+1}+1)}\over {2^{d_1}-1}}$  \\
\hline
$-2^{{m_1}+{d_1}}$ &   $2^{{n_1}-1}+2^{{m_1}+{d_1}-1}$        & ${(2^{{m_1}-{d_1}}-1)(2^{n_1}-1)}\over {2^{2{d_1}}-1} $   \\
\hline
$0$ &   $2^{{n_1}-1}$       & $ 2^{{m_1}-{d_1}}(2^{n_1}-1)$  \\
\hline
$2^{n_1}$ &   $0$       & $1$  \\
\hline
\end{tabular}
\end{center}
 \end{table}
\item For the case ${d_1'}=2{d_1}$ (the table at the top of next page),

 \begin{table*}[htbp]
 \begin{center}
 \begin{tabular}{|l|l|l|}
\hline
$value$  &   $weight$        & $multiplicity$   \\
\hline
 $-2^{m_1}$ &   $2^{{n_1}-1}+2^{{m_1}-1}$       & ${{2^{3{d_1}}(2^{m_1}-1)(2^{n_1}-2^{{n_1}-2{d_1}}-2^{{n_1}-3{d_1}}+2^{m_1}-2^{{m_1}-{d_1}}+1)}\over {(2^{d_1}+1)(2^{2{d_1}}-1)}}$  \\
\hline
$2^{{m_1}+{d_1}}$ &   $2^{{n_1}-1}-2^{{m_1}+{d_1}-1}$        & ${{2^{{d_1}}(2^{n_1}-1)(2^{m_1}+2^{{m_1}-{d_1}}+2^{{m_1}-2{d_1}}+1)}\over {(2^{d_1}+1)^2}}$  \\
\hline
$-2^{{m_1}+2{d_1}}$ &   $2^{{n_1}-1}-2^{{m_1}+2{d_1}-1}$        & ${{(2^{{m_1}-{d_1}}-1)(2^{n_1}-1)}\over {(2^{d_1}+1)(2^{2{d_1}}-1)}}$  \\
\hline
$2^{m_1}$   &   $0$        & $1$   \\
\hline
\end{tabular}
\end{center}
 \end{table*}
\end{enumerate}
 \end{lemma}

According to the possible weights of the codewords $c(\alpha,\beta)$ in Lemma \ref{CCC221}, we have
\begin{lemma} \label{ccc222}
The cyclic code $\mathcal{C}_{i,{m\over 2}}$ with idempotent $\theta_{l_i}^*+\theta_{l_{m\over 2}}^*$ can not have only three possible nonzero weights ${2^{m-1}}, {2^{m-1}+2^t}$ and ${2^{m-1}-2^t}$.
\end{lemma}

\begin{lemma}\label{FT001}
There are the following results about the exponential sum $T(\alpha,\beta)$
\[
\left\{
\begin{array}{lll}
\sum\limits_{\alpha,\beta \in \mbox{F}_q}T(\alpha,\beta) &=&2^{2m} \\
\sum\limits_{\alpha,\beta \in \mbox{F}_q}T(\alpha,\beta)^2 &=&2^{3m}.
\end{array}
\right.
\]

\end{lemma}

\begin{proof}
Exchanging the order of summation
\[
\begin{array}{lll}
&&\sum\limits_{\alpha,\beta \in \mbox{F}_q}T(\alpha,\beta) \\
& =  & \sum\limits_{\alpha,\beta \in \mbox{F}_q}\sum\limits_{x \in \mbox{F}_q}(-1)^{\mbox{Tr}\left(\alpha x^{2^i+1} + \beta x^{2^j+1}\right)}   \\
& = &  \sum\limits_{x\in \mbox{F}_q}\sum\limits_{\alpha \in \mbox{F}_q}(-1)^{\mbox{Tr}\left(\alpha x^{2^i+1}\right)}\sum\limits_{\beta \in \mbox{F}_q}(-1)^{\mbox{Tr}\left(\beta x^{2^j+1}\right)}  \\
& = &  q\cdot \sum\limits_{\stackrel{\alpha \in \mbox{F}_q}{x=0}}(-1)^{\mbox{Tr}\left(\alpha x^{2^i+1}\right)}=2^{2m};
\end{array}
\]
\begin{eqnarray*}
&&\sum\limits_{\alpha,\beta \in \mbox{F}_q}T(\alpha,\beta)^2\\
 & = & \sum\limits_{x,y\in \mbox{F}_q}\sum\limits_{\alpha \in \mbox{F}_q}(-1)^{\mbox{Tr}\left(\alpha \left( x^{2^i+1}+y^{2^i+1}\right)\right)}\\
 && \sum\limits_{\beta \in \mbox{F}_q}(-1)^{\mbox{Tr}\left(\beta \left( x^{2^j+1}+y^{2^j+1}\right)\right)} \nonumber\\
 & = & M_2\cdot 2^{2m}, \nonumber
\end{eqnarray*}
where $M_2$ is the number of solutions to the equation
\begin{equation}\label{ES2}
\left\{
\begin{array}{lll}
x^{2^i+1}+y^{2^i+1}&=& 0   \\
x^{2^j+1}+y^{2^j+1}&=&0.
\end{array}
\right.
\end{equation}
For any given $x\in \mbox{F}_q$, since $\mbox{gcd}(2^i+1,2^m-1)=1$, 
there is a unique $y\in \mbox{F}_q$ which satisfies the first one of the above equation system (\ref{ES2}), and thus $y=x$. 
Therefore $M_2=q=2^m$ and the result is obtained.
\end{proof}

\begin{lemma} \label{TM001}
The number of solutions of the following polynomial euqaiton system
\[
\left\{
\begin{array}{lll}
 x^{2^i+1}   +   y^{2^i+1}   +  z^{2^i+1} & =&  0\\
 x^{2^j+1}   +   y^{2^j+1}   +  z^{2^j+1}  &=&  0,
\end{array}
\right.
\]
is
\[
\begin{array}{lll}
 M_3 &=& (2^m-1)\\
  &&\left(2^{\mbox{gcd}(|i-j|,m)} + 2^{\mbox{gcd}(i+j,m)} - 2^{\mbox{gcd}(|i-j|,i+j,m)}\right)\\
  &+&2^m.
\end{array}
\]
\end{lemma}

\begin{proof}
Here, only the situation $i> j$ is considered.
 Divide both sides of the two equations by $z^{2^i+1}$ and $z^{2^j+1}$ respectively, and then after simplification they become:

 \begin{equation} \label{HERC}
\left\{
\begin{array}{lll}
 x^{2^i+1}   +   y^{2^i+1}   + 1  & =&  0\\
 x^{2^j+1}   +   y^{2^j+1}   + 1  &= & 0.
\end{array}
\right.
\end{equation}
Canceling $y$ we have ${\left(x^{2^i+1}+1 \right)}^{2^j+1}={\left(x^{2^j+1}+1 \right)}^{2^i+1}$ which is equivalent to

\[
{\left( x^{2^i}+x^{2^j}\right) }{\left( {x^{2^{i+j}}+x}\right) }={\left( {x^{2^{i-j}}+x}\right)}^{2^j}{\left( {x^{2^{i+j}}+x}\right)}=0.
\]
Therefore $x^{2^{i-j}}=x $  or  $x^{2^{i+j}}=x $, and let's consider them separately.

\noindent
Case I: $x^{2^{i-j}}=x $.\quad
 Set $k_1=\mbox{gcd}(i-j,m)$ and $q_1=2^{k_1}$, then $\mbox{F}_{q_1}=\mbox{F}_{2^{k_1}}$. 
 For any $x\in \mbox{F}_{q_1}$, since  $\mbox{gcd}(2^i+1,2^m-1)=1$, there is a unique element $y \in \mbox{F}_q$, such that $ x^{2^i+1} + y^{2^i+1}+ 1 = 0$.
Shift elements of the last equation, then take the exponential power $2^{i-j}$:
\begin{equation}\label{CTES}
\left( y^{2^i+1} \right) ^{2^{i-j}}   =  \left( x^{2^i+1} +1 \right) ^{2^{i-j}} = \left(x^{2^{i-j}}\right)^{2^i+1} +1 =  x^{2^i+1} +1.
\end{equation}
In the last step, we have used the fact that $x\in \mbox{F}_{q_1} \subset \mbox{F}_{2^{i-j}}$ that is $x^{2^{i-j}}=x$.
Comparing the left most and right most sides of equation (\ref{CTES}) to the first one of (\ref{HERC}), $\left( y^{2^i+1} \right) ^{2^{i-j}}=  y^{2^i+1}$.
That is
\begin{equation} \label{DETY}
\left( y^{2^{i-j}}\right)^{2^{i}+1} = y^{2^i+1}.
\end{equation}
Using again the fact that $\mbox{gcd}(2^i+1,2^m-1)=1$, equation (\ref{DETY}) implies that $y^{2^{i-j}}=y$, so $y \in \mbox{F}_{q_1}$.

  For $x, y \in \mbox{F}_{q_1}$, take the exponential power $2^{i-j}$ of the second equation of (\ref{HERC}):
 \begin{eqnarray}
  \left( x^{2^j+1}   +   y^{2^j+1}   + 1\right)^{2^{i-j}} &=& x^{2^{i}+2^{i-j}} +y^{2^{i}+2^{i-j}} +1 \nonumber\\
   &=&x^{2^{i}}\cdot x^{2^{i-j}}+ y^{2^{i}}\cdot y^{2^{i-j}} +1 \nonumber\\
  &=&  x^{2^i+1}   +   y^{2^i+1}   + 1, \nonumber
 \end{eqnarray}
 which implies that the two equations of (\ref{HERC}) are equivalent. 
So, the number of solutions $(x,y)$ of (\ref{HERC}) in $\mbox{F}_{q_1}$ is $N_1' = q_1$. 

  Case II: $x^{2^{i+j}}=x $.\quad Set $k_2=\mbox{gcd}( {i+j}, m)$, $q_2=2^{k_2}$ and $\mbox{F}_{q_2}=\mbox{F}_{2^{ k_2}}$. 
Let $N_2'$ be the number of $(x,y) \in \mbox{F}_{q_2}^2$ satisfying (\ref{HERC}), then similarly we have $N_2' = q_2$. 

For the joint of the solution sets of the two cases, set $k_3=\mbox{gcd}(i-j,i+j,m)$, $q_3=2^{ k_3}$ and $\mbox{F}_{q_3}=\mbox{F}_{2^{k_3}}$. 
  Then the number of $(x,y) \in \mbox{F}_{q_3}^2$ satisfying (\ref{HERC}) is $N_3' = q_3$. 

Combing above, the number of $(x,y) \in F_{q }^2$ satisfying (\ref{HERC}) is $N' = N_1'+N_2'-N_3'$. 
Thus $M_3=(q-1)N'+M_2$, and the result of the lemma is obtained.
\end{proof}

\begin{corollary}\label{TM002}
There is the following result about the exponential sum $T(\alpha,\beta)$
\[
\begin{array}{lll}
\sum\limits_{\alpha,\beta \in \mbox{F}_q}T(\alpha,\beta)^3 &=&2^{2m}M_3.
\end{array}
\]
\end{corollary}

Let $f(x)=\alpha x^{2^i+1} + \beta x^{2^j+1}$, where $(\alpha,\beta)\in \mbox{F}_q^2\backslash\{(0,0)\}$.
According to the relation between the weight of a codeword and corresponding exponential sum (\ref{wt}),
 assume that $T(\alpha,\beta)$ takes only three possible values $\pm 2^{m\over 2}, 0$.
For $\varepsilon = \pm 1$, define $N_{\varepsilon,0} = \{  (\alpha, \beta) \in \mbox{F}_q \times \mbox{F}_q\backslash \{(0,0)\}|T(\alpha,\beta) =\varepsilon 2^{m\over 2}\}$,
and $n_{\varepsilon,0} =|N_{\varepsilon,0}|$; $N_{ 0} = \{  (\alpha, \beta) \in \mbox{F}_q \times \mbox{F}_q\backslash \{(0,0)\} |T(\alpha,\beta) =0\}$
and $n_0 =|N_{0}|$.

\begin{lemma}\label{FT002}
Under above specifications, 
\[
\begin{array}{lll}
 & n_0  = 2^m-1, n_{1,0} = {1\over 2}(2^{2m}+2^{{3\over 2}m}-2^m-2^{m\over 2}), \\
 &\mbox{and} \  n_{-1,0}={1\over 2}(2^{2m}-2^{{3\over 2}m}-2^m+2^{m\over 2}).
  \end{array}
  \]
\end{lemma}

\begin{proof}
Substituting the notations to Lemma \ref{FT001} 
\[
\left\{
\begin{array}{lll}
n_0+n_{1,0} + n_{-1,0} &=& 2^{2m} -1 \\
2^m + 2^{m\over 2}\cdot n_{1,0} - 2^{m\over 2}\cdot n_{-1,0} &=& 2^{2m} \\
2^{2m} + 2^{m }\cdot n_{1,0} + 2^{m }\cdot n_{-1,0}& =& 2^{3m},
\end{array}
\right.
\]
note that $T(\alpha = \beta =0) =q=2^m$.
\end{proof}

Parallel to Lemma \ref{ccc222}, the following lemma is used to characterize the weight distribution of the cyclic code $\mathcal{C}_{i,j}$ where $1\leq i\not= j\leq t$ and $\mbox{gcd}(2^i+1,2^m-1)=1$.

\begin{lemma}\label{EM222}
Assume $m=2t+2$ is not a power of $2$. Then the cyclic code $\mathcal{C}_{i,j}$  with idempotent $\theta_{l_i}^* + \theta_{l_j}^*$ can not have only three possible nonzero weights $ 2^{m-1}, 2^{m-1}-2^{{m\over 2} -1} $ and
$2^{m-1}+2^{{m\over 2} -1}$.
\end{lemma}

\begin{IEEEproof}
If $\mathcal{C}_{i,j}$ has only those three nonzero weigths, from Corollary \ref{TM002} and Lemma \ref{FT002},
\[
\begin{array}{lll}
\sum\limits_{\alpha,\beta \in \mbox{F}_q}T(\alpha,\beta)^3  = 2^{3m} + 2^{{3m}\over 2}\cdot n_{1,0} - 2^{{3m}\over 2}\cdot n_{-1,0}  =2^{2m}M_3.
\end{array}
\]
That is $M_3= 2^m+2^{m} -1$.
Thus,
$ 2^{\mbox{gcd}(|i-j|,m)} + 2^{\mbox{gcd}(i+j,m)} - 2^{\mbox{gcd}(|i-j|,i+j,m)} =1$ which is impossible.
\end{IEEEproof}


 Lemma \ref{ccl} is a known result, and stated without proof. 
\begin{lemma} \label{ccl}
The irreducible cyclic code $\mathcal{C}_0$ with primitive idempotent $\theta_1^*$ has dimension $m$ and only one nonzero weight $2^{m-1}$.
\end{lemma}

\begin{corollary} \label{ECC222}
Assume $m= 2 t+2$ is not a power of $2$. Then in the process of selecting the optimum cyclic subcode chain of $\mathcal{RM}(2,m)^*$ under standard $II$, the first three primitive idempotents to be selected are $\theta_0,\theta_1^*$ and $\theta_{l_{m\over 2}}^*$. Then (\ref{OPECC}) is an optimum cyclic subcode chain.
\end{corollary}
\begin{IEEEproof}
In the selection process, it is easy to see that the first primitive idempotent is $\theta_0$, and the resulting cyclic code has weight $2^m-1$.

According to Lemma \ref{PRIMC} and Lemma \ref{ccl}, we see that the second primitive idempotent to be selected might be $\theta_1^*$ or $\theta_{l_i}^*$, where $1\leq i \leq t$ is a positive integer that satisfies $\mbox{gcd}(2^i+1,2^m-1)=1$. Consider the two cases separately.
\begin{enumerate}
\item
If the second primitive idempotent selected is $\theta_{l_i}^*$. Then from Lemma \ref{ccc222} and Lemma \ref{EM222} we find that, no matter which primitive idempotent of the form $\theta_{l_j}^*(j \not= i, 1\leq j \leq t+1)$ is selected in the third step, the corresponding cyclic subcode $\mathcal{C}_{i,j}'$ with idempotent $\theta_0+\theta_{l_i}^*+\theta_{l_j}^*$ has minimum distance less than $2^{m-1}-2^t-1$.
\item
If the second primitive idempotent selected is $\theta_1^*$, considering Theorem \ref{even22}, the third cyclic subcode with idempotent $\theta_0+\theta_{1}^*+\theta_{m\over 2}^*$ has minimum distance $2^{m-1}-2^t-1$.
\end{enumerate}

Comparing the above two cases, 
the result follows from Theorem \ref{even22}.
\end{IEEEproof}

 Combining Corollary \ref{EC222} and Corollary \ref{ECC222}, the main result of this section follows in Theorem \ref{even222} by extending Theorem \ref{even22}.




 \begin{theorem} \label{even222}
Let $m=2t+2$ where $t\geq 1$. Then for the punctured second-order Reed-Muller code $\mathcal{RM}(2,m)^*$, the distance profile of the cyclic subcode chain in Corollary \ref{evenc1} is optimum under Standard II.
 \end{theorem}

\begin{example}
Set $m=6=2\cdot 2+2$, i.e. $t=2$.
The optimum cyclic subcode chain given in Theorem \ref{even222} can be constructed as follows. Note that there are five primitive idempotents here $\theta_0 $, $\theta_1^*$,  $\theta_3^*$,  $  \theta_5^*$ and  $\theta_9^*$.
\begin{itemize}
\item
The minimum distance of the cyclic code with idempotent $\theta_0$ is $63$, and it is chosen  as the first cyclic subcode of the chain.
\item
The minimum distances of the cyclic subcodes with idempotents $\theta_0 + \theta_1^*$,  $\theta_0 + \theta_3^*$,  $\theta_0 + \theta_5^*$ and  $\theta_0 + \theta_9^*$ are $31, 24, 31$ and $27$ respectively. There are two choices for us: $\theta_0 + \theta_1^*$ or $\theta_0 + \theta_5^*$, which will be suggested later.
\item
The minimum distances of the cyclic subcodes with idempotents $\theta_0 + \theta_1^* + \theta_3^*$,  $\theta_0 + \theta_1^* + \theta_5^*$,  $\theta_0 + \theta_1^* + \theta_9^*$ and $\theta_0 + \theta_5^* + \theta_3^*$, $\theta_0 + \theta_5^* + \theta_9^*$ are $23, 23, 27$ and $24, 23$.  So, in this step  $\theta_0 + \theta_1^* + \theta_9^*$ is selected, and then in last step $\theta_0 + \theta_1^*$ is selected.
\item
The minimum distances of the cyclic subcodes with idempotents  $\theta_0 + \theta_1^* + \theta_9^* + \theta_3^*$ and  $\theta_0 + \theta_1^* + \theta_9^*+\theta_5^*$  are $15$ and $23$ respectively.
Select  $\theta_0 + \theta_1^* + \theta_9^* + \theta_5^*$ in this step.
 \item
 Finally the minimum distance of the cyclic code with idempotent $\theta_0 + \theta_1^* + \theta_9^* + \theta_5^*+\theta_3^*$, i.e $\mathcal{RM}(2,m)^*$, is $15$.
\end{itemize}
Therefore,  the ODPC-II$^{inv}$ of the punctured second-order Reed-Muller code $\mathcal{RM}(2,6)^*$ is $d_{\tau_{0}}=15,\, d_{\tau_{1}}=23,\, d_{\tau_{2}}=27,\, d_{\tau_{3}}=31,\, d_{\tau_{4}}=63$.

\end{example}

\section{Suboptimums with respect to ODPC-I$^{inv}$ of $\mathcal{RM}(2,m)^*$ and one optimum} \label{secV}

 In this section, the cyclic subcode chain of the punctured second-order Reed-Muller code $\mathcal{RM}(2,m)^*$ is studied under Standard I. Proposition \ref{MNESI} of Subsection \ref{secV.I} gives a suboptimum result with respect to ODPC-I$^{inv}$ for the case $m=2t+1$, which considers almost all the subcode chain classes respectively. Proposition \ref{MESI} of Subsection \ref{secV.II} concerns the case where $m=2t+2$ for almost  half of the subcode chain classes, and Corollary \ref{007} emphasizes that in fact the optimum result can be obtained when $m$ is a power of $2$.

\subsection{The Case of $m=2t+1$ } \label{secV.I}

From Lemma \ref{th1}, the length of the cyclic subcode chains is $\lambda = t+2$, the number of the cyclic subcode chains is $\lambda ! = (t+2)!$. The number of the chains in each class is $\mu = (t+1)! \cdot 1! =(t+1)!$,
the number of the classes is $t+2$. For the study of the ODPC-I$^{inv}$, consider the dimension profile
\begin{eqnarray} \label{001}
(t+1)m+1, \ldots, um+1, \ldots, 2m, m,
\end{eqnarray}
where $2\leq u \leq t$.

\begin{proposition} \label{MNESI}
Let $m=2t+1$ where $t\geq 2$. For the code $\mathcal{RM}(2,m)^*$, consider Standard I with dimension profile (\ref{001}). If we are requiring that, the cyclic subcode $\mathcal{C}_0$ or equivalently the primitive idempotent $\theta_1^*$ is selected first, the cyclic subcode chain obtained by adding the primitive idempotents one by one in the following order is optimum:
    \[
    \theta_1^*, \theta_{l_{t}}^*,\ldots,  \theta_{l_{t-u+2}}^*, \theta_0, \theta_{l_{t-u+1}}^*, \ldots,\theta_{l_{1}}^*.
   \]
  And the distance profile is
   \[
   \begin{array}{l}
   d_{\tau_{v}}=2^{2t}-2^{2t-{v}-1}-1 (0\leq v\leq t-u+1) \\
   d_{\tau_{v}}=2^{2t}-2^{2t-v} (t-u+2\leq v\leq t)\\
   d_{\tau_{t+1}} = 2^{m-1}=2^{2t},
   \end{array}
   \]
  where $\theta_0$ is selected to be added in the $(u+1)$th order.
\end{proposition}

\begin{example}
For $t=2$, that is $m=5, n=31, u=2$. Proposition \ref{MNESI} provides a distance profile $d_{\tau_0}=7,\, d_{\tau_1}=11,\, d_{\tau_2}=12,\, d_{\tau_3}=16$ with dimension profile $16,\, 11,\, 10,\, 5$. The generated cyclic codes [31,11,11], [31, 10, 12] and [31, 5, 16] are optimal \cite{G}.
\end{example}

\subsection{The Case of $m=2t+2$} \label{secV.II}

  In this case, the length of the cyclic subcode chains is $\lambda = t+3$, and the number of the cyclic subcode chains is $\lambda ! = (t+3)!$. The number of chains in each class is $\mu = (t+1)! \cdot 1!\cdot 1! =(t+1)!$,
and the number of the classes is $(t+3)(t+2)$. 
 In Proposition \ref{MESI}, a suboptimum ODPC-I$^{inv}$ of $\mathcal{RM}(2,m)^*$ is presented, with corresponding dimension profile
\begin{equation}\label{002}
  \begin{array}{lll}
 & (t+1)m+{m\over 2}+1, \ldots, (j-1)m+{m\over 2}+1,  \ldots, \\
  & im+{m\over 2}, \ldots, 2m, m,
 \end{array}
 \end{equation}
where $2\leq i <j\leq t+1$.

\begin{proposition} \label{MESI}
Let $m=2t+2$ where $t\geq 2$. For the code $\mathcal{RM}(2,m)^*$, consider Standard I with dimension profile (\ref{002}). If we are requiring that, the cyclic subcode $\mathcal{C}_0$ or equivalently the primitive idempotent $\theta_1^*$ is selected first, the cyclic subcode chain obtained by adding the primitive idempotents one by one in the following order is optimum:
   \[
   \begin{array}{lll}
& \theta_1^*, \theta_{l_t}^*, \theta_{l_{t-1}}^*,\ldots, \theta_{l_{t-i+2}}^*,\theta_{l_{m\over 2}}^*, \theta_{l_{t-i+1}}^*,\ldots, \\
  &  \theta_{l_{t-j+3}}^*,\theta_{0} , \theta_{l_{t-j+2}}^*,\ldots,\theta_{l_{1}}^*.
    \end{array}
  \]
   And the distance profile is
   \[
   \begin{array}{l}
    d_{\tau_v}=2^{2t+1}-2^{2t-v}-1 (0\leq v\leq t-j+2),   \\
      d_{\tau_v}=2^{2t+1}-2^{2t-v+1} (t-j+3\leq v\leq t-i+2),   \\
     d_{\tau_v}=2^{2t+1}-2^{2t-v+2} (t-i+3\leq v\leq t+1),  \\
       d_{\tau_{t+2}} =  2^{m-1}=2^{2t+1},
       \end{array}
       \]
  where $\theta_{l_{m\over 2}}^*$ is selected in the $(i+1)$th order, and $\theta_0$ is selected in the  $(j+1)$th order.
  \end{proposition}

\begin{example}
For $t=2$, that is $m=6, n=63, i=2, j=3$. Proposition \ref{MESI} provides a distance profile $d_{\tau_0}=15,\, d_{\tau_1}=23,\, d_{\tau_2}=24,\, d_{\tau_3}=24,\, d_{\tau_4}=32$ with dimension profile $22,\, 16,\, 15,\, 12,\, 6$. The generated cyclic codes [63,16,23], [63, 15, 24], [63, 12, 24] and [63, 6, 32] are almost optimal \cite{G}.

\end{example}

\begin{corollary}\label{007}
In Proposition \ref{MESI}, if $m=2^s$ ($s\geq 2$), from Corollary \ref{EE22} we do not require the preassumption that the primitive idempotent $\theta_1^*$ is the first to be selected, since $\theta_1^*$ corresponds to the unique nontrivial irreducible cyclic code with minimum distance $2^{m-1}$. 
\end{corollary}

\section{Conclusion} \label{secVII}

The optimum distance profile serves as a new research field in coding theory. It has been investigated for the generalized Reed-Solomon code, the Golay code, the first-order Reed-Muller code, the second-order Reed-Muller code, and some other codes in \cite{HL}, \cite{HL1} and \cite{CL}. Known results on the distance profile of the linear codes can be applied to construct polar codes with good polarizing exponents \cite{SE}.
Rather than the general linear codes, this paper studies cyclic codes and their cyclic subcode chains because of easy encoding and more algebraic structures. 

For the punctured second-order Reed-Muller code $\mathcal{RM}(2,m)^*$, suboptimum results (optimum under certain requirement) of two standards about the ODPC are presented in Theorem \ref{th4}, Theorem \ref{even22}, Proposition \ref{MNESI} and Proposition \ref{MESI},
the requirement of which is common such that only the primitive idempotent $\theta_1^*$ is fixed early. And the results
deduce the ODPCs for the case of $m=2t+2$, see Theorem \ref{even222} and Corollary \ref{007}.

\section*{Acknowledgments}

This cooperative work was mainly finished during a visit in Institute of Network Coding, the Chinese University of Hong Kong, Hong Kong, P. R. China. Thanks a lot for the warm reception of Professor Raymond Yeung and General Manager HO, Chi-lam Alfred.

\end{document}